# Atmospheric Modelling for the Removal of Telluric Features from Infrared Planetary Spectra

Daniel V. Cotton[*], Jeremy Bailey and Lucyna Kedziora-Chudczer

*School of Physics, University of New South Wales, NSW 2052, Australia*

[*] E-mail: d.cotton@unsw.edu.au

**ABSTRACT**

The effects of telluric absorption on infrared spectra present a problem for the observer. Strong molecular absorptions from species whose concentrations vary with time can be particularly challenging to remove precisely. Yet removing these effects is key to accurately determining the composition of many astronomical objects – planetary atmospheres in particular. Here we present a method for removing telluric effects based on a modelling approach. The method relies only on observations usually made by the planetary astronomer, and so is directly comparable with current techniques. We use the modelling approach to process observations made of Jupiter, and Saturn's moon Titan and compare the results with those of the standard telluric division technique, finding the modelling approach to have distinct advantages even in conditions regarded as ideal for telluric division.

## 1 INTRODUCTION

Ground based observations, particularly in the infrared, are usually limited to spectral windows between the strong absorptions of molecules in the Earth's atmosphere. The strongest absorption lines, belong to some of the species that are most abundant in the Earth's atmosphere; namely water vapour ($H_2O$), carbon dioxide ($CO_2$), methane ($CH_4$) and nitrous oxide ($N_2O$) (Seifahrt et al. 2010, Bailey et al. 2007).

When observing solar system- and exo- planetary atmospheres in particular these same molecular absorptions are frequently of interest. For instance, $H_2O$, $CH_4$, $CO_2$ and carbon monoxide (CO) have all been detected in the atmospheres of hot Jupiters (Swain et al. 2009) with space- and ground-based instruments. Indeed $H_2O$ is predicted to be among the most abundant molecular species, after hydrogen ($H_2$), in close-in giant planets (Tinetti et al. 2007). $CO_2$ signatures dominate the Martian and Venusian infrared spectra (Encrenaz 2008), while the giant planets largely have their spectra dominated by $CH_4$ absorption; with $CO_2$, CO and $H_2O$ being only minor constituents – but no less of interest (Encrenaz 2008).

To be able to get precise determinations of molecular abundances from planetary observations with ground based telescopes, the effects of the Earth's atmosphere need to be accurately removed. This task is complicated by temporal and spatial variations of the telluric absorbing species. It is well known that $CO_2$ levels are rising worldwide (Raupach et al. 2007), but overlayed with this trend are various seasonal and other short-term influences (Buemann et al. 2007) that make accurately predicting its line-of-sight concentration at a particular time difficult. Variations in $H_2O$ and ozone ($O_3$) concentrations can be even more rapid (Bailer-Jones & Lamm 2003).

The method normally employed for removing the effects of the telluric atmosphere – called telluric division – involves taking the spectrum of interest and dividing it by a telluric standard, which could be a star without strong features in its spectrum. Telluric absorption standards for spectral observations of solar system planets are usually solar type stars viewed as much as possible through the identical slab of the Earth's atmosphere in a spatial and temporal sense (Seifahrt et al. 2010, Bailey et al. 2007). The lists of telluric standards for near-IR observations of southern objects are given by McGregor (1994), Carter (1990), Bouchet, Manfroid, & Schmider (1991) and a selection of G-type standard stars has been compiled by Allen and Cragg (1983). The best telluric standard would be featureless across the region of interest (e.g. spectral types mid B- to late A-type approximate this in regions between their strong hydrogen lines); or a standard with well characterised features (e.g. solar-like G type stars paired with high resolution Solar data) (Seifahrt et al. 2010). However, "featureless" standards do not exist and atmospheric absorption can vary on the timescale of minutes.

Typical correction of telluric features is an iterative procedure that relies on gradual adjustment of the wavelength and intensity scale to match the object and calibration spectrum before division. However this can be a time-consuming process and the quality of the division is decided upon by the subjective, "by-eye" judgement of the observer. Data reduction tasks such as TELLURIC in the IRAF[1] package improve this situation by providing an automated procedure for subtraction of atmospheric features, where the scaling of the telluric standard spectrum follows a simple Beer's law (Beer 1852) approximation for the curve of growth of telluric lines. The correction for differences in the zenith angle between the observing target and the telluric spectrum is made by assuming linear extinction with air-mass. This is a reasonable assumption to make at visible wavelengths where Rayleigh and aerosol scattering dominate, but in the near infra-red, strong saturated molecular absorptions result in extinction being decidedly non-linear (Bailey et al. 2007).

In addition, any stellar features in the telluric standard cannot be automatically removed leaving spurious artefacts in the target spectrum. In observations of solar system planets there is a considerable effort to find solar type telluric standard stars to accurately remove the Solar lines reflected by the planet's atmosphere. However in high-resolution spectroscopy the difference in Doppler shift between the telluric standard star and the observed planet[2] has to be accounted for independently.

A more fundamental problem with telluric division is encountered when the spectra are taken with all but the highest resolution instruments. To fully resolve individual telluric absorption lines in the near-infrared can require a resolving power of R ($\lambda/\Delta\lambda$) > 100,000 (Bailey et al. 2007, Seifahrt et al. 2010). However, a low resolution telluric spectrum cannot properly represent the complex high resolution structure that is unresolved, and therefore cannot properly correct for it. Simulations (Bailey et al. 2007) showed that large errors can be introduced, particularly in the case where the object being observed has the same absorption species as the Earth's atmosphere (e.g. observations of $CO_2$ in the Martian atmosphere).

---

[1] IRAF is distributed by the National Optical Astronomy Observatory, which is operated by the Association of Universities for Research in Astronomy (AURA) under cooperative agreement with the National Science Foundation.

[2] The target object is typically anything with an atmosphere from a moon such as Titan to a solar system- or exo-planet to a brown dwarf. To simplify the language of the paper we will refer to all of these as planets.

In this paper we describe an alternative approach to correction of the atmospheric telluric absorption. We derive a spectrum of the telluric absorption lines based on a model of the Earth's atmosphere at a time and position of the target observation. The telluric transmission spectrum is modelled by using VSTAR, line-by-line radiative transfer software. Similar models used in the past were either empirical (where a catalogue was derived from measurements of the atmosphere) (Lundström *et al.* 1991) or theoretical (Bailey et al. 2007, Lundström et al. 1991, Seifahrt et al. 2010, Widemann *et al.* 1994) in nature. The availability of extensive molecular absorption line databases that are continually being improved upon (e.g. HITRAN) and well determined atmospheric profiles for the major components of the telluric atmosphere make theoretical models increasingly accurate and as such are favoured over an empirical approach.

The most sophisticated of the modelling approaches is forward modelling where the telluric absorption lines and the absorption features of the target source are simultaneously modelled and both are iteratively altered to fit the data (Chamberlain *et al.* 2006, Bailey et al. 2007). A forward modelling technique allows the removal of telluric absorption features with all lines resolved to arbitrarily high resolution, before an instrumental response correction is applied. Such an approach is only possible if sufficient information is available about the structure and constituents present in the atmosphere of the target object.

If the spectrum of the object being observed is sufficiently different to that of the telluric atmosphere, it is possible to fit the atmospheric model directly to the data (Bailey et al. 2007, Lallement *et al.* 1993); otherwise separate measurements of the telluric atmosphere are needed. One method is to use information gathered by meteorology and forecasting agencies to determine the telluric atmospheric transmission (Seifahrt et al. 2010); this approach has the advantage that no additional measurements are required and also allows for predictions of signal-to-noise to be made ahead of measurements based on forecasts; the disadvantage is that no *in situ* measurement of the atmosphere along the line-of-sight is made, and that the astronomer is relying upon information being gathered elsewhere being relevant for the observatory. In contrast our method uses observations of a telluric standard to determine the state of the atmosphere, by fitting a model of the telluric atmospheric transmission to match the observed spectrum.

All of the modelling approaches for removing telluric lines described have another important advantage over telluric division – they allow for improved wavelength calibration. A precise wavelength calibration is important for high resolution spectra, and for modelling planetary atmosphere spectra. Rare gas (He, Ne, Ar, Xe, Kr) emission line lamps are usually used for wavelength calibration, but these provide only a sparse line coverage in many parts of the near-infrared spectrum (Seifahrt et al. 2010). By contrast there are a lot of telluric absorption lines in this region, and with modelling these can be used for wavelength calibration.

VSTAR (Versatile Software for Transfer of Atmospheric Radiation) is a line-by-line radiative transfer code that can be used to model atmospheres of everything from the Earth and other terrestrial planets and moons to giant planets, brown dwarfs and cool stars (Bailey & Kedziora-Chudczer 2012). VSTAR is particularly appropriate to integrate into code for the removal of telluric lines from astronomical spectra since it's flexibility makes it suitable to a forward modelling approach for a wide variety of target objects.

This paper introduces ATMOF (ATMOspheric Fitting) code and procedures, that utilise VSTAR, that we have developed and applied to observations of Titan and Jupiter. The methodology applied is that of using a standard star observation to determine the state of the atmosphere and then apply it to the science target observations; this is valuable in its own right, but is also a stepping-stone to full forward modelling.

To demonstrate advantages of using the ATMOF code, we apply it to high resolution spectra of Jupiter and Saturn's moon Titan, obtained with the GNIRS instrument at the Gemini North telescope. The spectra are at wavelengths of ~2 μm, where there are strong telluric $CO_2$ absorption bands that need to be removed. This region of the spectrum is interesting because it is one of the spectroscopic "windows" between strong methane absorptions that allows us to see deeply into the atmosphere of Jupiter, and to the surface of Titan.

We compare our modelling approach with the standard telluric division method of scaling and division of the spectrum by the telluric standard star spectrum.

In Section 2 we present our Gemini observations and common data reduction method. Next we describe the ATMOF modelling approach (Section 3 ). In section 4 we compare both methods of telluric correction the traditional division and the ATMOF modelling for both Jupiter and Titan. Finally we discuss applications of the ATMOF method to high-resolution spectroscopic observations in near infrared and also its potential for use in radial velocity searched for exoplanets (Section 5 ).

## 2 OBSERVATIONS AND STANDARD DATA REDUCTION

### 2.1 Observations

Jupiter's and Titan's long-slit spectra were acquired at the highest resolving power (R ~ 18000) available with the GNIRS instrument at the Gemini North 8 m telescope, as part of the program to search for trace atmospheric molecules and measure the isotope ratio of deuterium to hydrogen using lines of the methane isotopologues (Kedziora-Chudczer *et al.* 2013). Here we present data obtained in the near-infrared K band centered on 2.03 μm for Titan and 2.04 μm for Jupiter respectively. The details of observing conditions and instrumental settings are presented in Table 1. The slit used was 0.1 arcsecond wide – much narrower than the seeing at the time of the observations, which was, for example, ~0.6 to 0.7 arcsecond on the night of the Titan observation – and was oriented to cover the extent of both objects in the direction as close as possible to the north-south axis.

Table 1. Observing conditions and observational architecture.

| Observing | Jupiter | Titan |
|---|---|---|
| Date | 27-Aug-11 | 31-May-11 |
| Time (UT) | 14:51:46 | 9:12:52 |
| GNIRS grating | 111/mm LB_G5534 | 111/mm LB_G5534 |
| Total integration time | 4 x 160 sec | 4 x 300 sec |
| Central wavelength | 2.04 μm | 2.03 μm |
| Mean Airmass | 1.011 | 1.431 |
| Humidity percentage | 7-9 | 69-71 |
| Telluric standard star | HIP 11053 | HIP 64345 |

Our observing strategy for Titan was to obtain four 300 sec exposures with the source moved between two slit positions (A and B) in an ABBA sequence. This mode of observing is more efficient as separate observations of sky are not needed. However for the highly extended planets like Jupiter this shortcut could not be applied and four additional 160 sec exposures of sky in the vicinity of the planet were obtained following pattern ABBAABBA (where A corresponds to sky and B planetary spectrum). Wavelength calibration lamps (Ar-Xe), flat-fields and telluric standards were observed immediately before or after the full cycle of exposures for each target object and sky.

## 2.2 Common Data reduction

Gemini IRAF scripts for GNIRS were used for the first stages of data reduction. After visual examination of 2D spectra and correcting of the vertical striping pattern due to variable bias that was apparent in some frames, all data frames were aligned, corrected for bias level, non-linearity and bad pixels on the CCD detector. Next the objects and standard stars frames were flat-fielded and wavelength calibrated. For the latter we used the interactive procedure to ensure correct identification of a small number of Argon and Xenon lines present in the spectral region between 2.0 and 2.06 μm.

In these high-resolution spectra the number of lines in the calibration lamp was insufficient to give a reliable wavelength calibration across the whole spectrum. An advantage of the modelling technique described here is that it enables an improved wavelength calibration using the telluric lines as a wavelength reference.

Sky subtraction was performed by combining and subtracting frames with the common target offset on the slit for the telluric standards and Titan. For Jupiter the combined sky frame was subtracted from the object frame. Finally dispersion correction was applied to our targets and the 1D straightened spectra were extracted in the region of the highest intensity along the 2D frame. This corresponds to the bright equatorial band for Jupiter. The extracted averaged spectrum of Titan is also centred on the equatorial region of the moon, however it encompasses a much larger latitudinal extent.

## 2.3 Telluric Division

Telluric division is the standard data reduction technique, to which we will be comparing the results of our modelling approach.

The extracted Titan and Jupiter spectra are dominated by the strong absorption of $CO_2$ and $H_2O$ bands present in the Earth's atmosphere. We used the traditional method of correcting this telluric absorption by dividing spectra of our targets by corresponding spectra of telluric standard stars. The recommended IRAF package routine, TELLURIC was used, and this allows scaling of the telluric standard spectrum and small wavelength shifts to be applied iteratively until the target spectrum appears to be free of telluric absorption. This technique worked fairly well for observations of Jupiter that were taken in good weather, low humidity and at a high elevation angle.

Observations of Titan were taken at a lower elevation angle in more humid conditions. Additional problem may have been less accurate wavelength calibration due to lack of the good argon and xenon lines at the longer wavelength end, while the Jupiter spectrum centred on the slightly longer wavelength did not have this problem. Telluric correction for Titan appeared to work well either at one or the other end of the spectrum when slightly different wavelength shifts were applied within the range of 0.1 Angstrom, that confirmed wavelength calibration limitations for this object.

## 3 MODELLING APPROACH

Our method can be broken up into two parts: (1) determining the state of the telluric atmosphere and the instrumental response, and (2) removing the affects of the atmosphere and instrumental response from the data. To do this, in addition to the observations already described in section 2.1, we require a telluric star reference spectrum, a model of the telluric atmosphere, and appropriate fitting software, as well as an ability to take account of instrumental response.

### 3.1.1 High resolution solar spectrum

We use a high resolution solar spectrum as a proxy for an observed standard star. The high resolution data comes from the Kitt Peak Solar Atlas (Wallace *et al.* 1996); it is provided freely online by the National Solar Observatory (NSO), in overlapping 25 $cm^{-1}$ segments of resolution ~0.01 $cm^{-1}$ (R = 300,000), that we then stitch together. This data set provides, as well as the directly observed solar spectra, data that has been separated into the telluric and purely solar components. It is the latter that is used to provide the solar spectrum for our modelling.

The stitching was done using an Integrated Development Language (IDL) based least squares fitting routine. A scaling factor was the only free parameter used to align the overlapping regions, the points overlapping denoting which being subsequently replaced by their averages. Where the Kitt Peak data is incomplete we replace the entire 25 $cm^{-1}$ segment by the lower resolution Kurucz synthetic solar spectrum[3]. The result is a neatly stitched spectrum, but one with a trend influenced by detector sensitivity, and different from that of the Kurucz synthetic solar spectrum.

We then use another IDL routine to fit both the Kitt Peak data and Kurucz's synthetic solar spectrum with polynomials, and then scale each Kitt Peak data point by the ratio of the two polynomials at that point.

The result is a solar spectrum of high resolution on an absolute flux scale (deriving from the Kurucz synthetic data). The section of the solar spectrum used for the observations described here is shown in Fig. 1. The Kitt Peak data is available for a much larger spectral range than presented here – from 1850 to 9000 cm$^{-1}$. We have processed much of that data to obtain high resolution flux calibrated solar data for the J, H and K regions of the infrared spectrum.

---

[3] Kurucz, R L. Obtained online 12th of January 2009. The solar irradiance by computation, http://kurucz.harvard.edu/papers/irradiance/solarirr.tab.

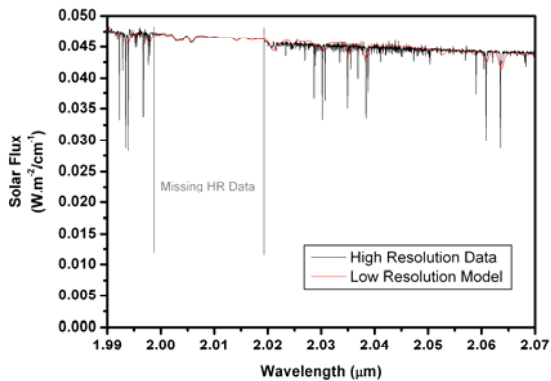

Fig. 1. The spectral standard derived from high resolution Kitt Peak Solar Atlas data, and the lower resolution Kurucz synthetic solar spectrum.

The effect of the missing high resolution data between ~2 and 2.02 μm on the standard star fits will be shown and discussed in section 4.

*3.1.2 VSTAR telluric atmosphere model*

Modelling the telluric atmosphere is a particularly simple matter in VSTAR, as a specific routine allowing the use of standard atmospheres for various locations and seasons on the Earth can be employed.

The telluric atmosphere model used is the Manu Kea (MK) profile built-in to VSTAR. This profile is based on the Mid Latitude Summer (MLS) profile but with the lower levels removed to match the height of the observatory. Molecular absorption line data are taken from the HITRAN 2008 database (Rothman *et al.* 2009). The model includes absorption from $CO_2$, $H_2O$, $O_2$, $CH_4$, CO, $O_3$ and $N_2O$; and Rayleigh scattering in air. The wavenumber grid is 0.01 cm$^{-1}$ (R ~500,000) with a range just beyond that of the data. An example of the transmission spectrum calculated at this resolution is given in Fig. 2.

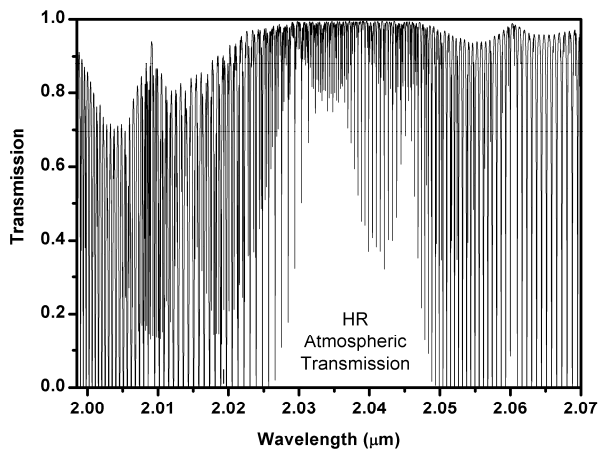

Fig. 2. An example of the high resolution atmospheric transmission spectrum calculated with VSTAR.

Transmission spectra for the atmosphere are derived using the line-by-line method. Each spectral line is individually modelled with its correct line shape at each layer in the altitudinal profile and the layer optical depths are added together to give the overall transmission for any zenith distance. To match observations, the $H_2O$ content was altered by scaling the entire altitudinal profile below the stratopause (65 km), and the $CO_2$ content by scaling the entire altitudinal profile; the scaling ratio being determined by ATMOF in accordance with procedures outlined below.

*3.1.3 ATMOF*

ATMOF is a modular fitting program coded in FORTRAN. Its role is to determine the state of the telluric atmosphere, and the current instrumental response by comparing a reference spectrum, augmented by a telluric transmission spectrum and instrumental response, with the observational data. ATMOF uses the Levenberg-Marquardt method for fitting (Press *et al.* 1992). The Levenberg-Marquardt algorithm is a standard nonlinear least-squares routine; it varies smoothly between the extremes of the Inverse-Hessian Method (used close to minimum) and the Steepest Descent Method (used far from minimum).

ATMOF is modular and uses a VSTAR model of the telluric atmosphere (section 3.1.2), which calculates an atmospheric transmission spectrum based on input parameters such as mixing ratios for $CO_2$ and $H_2O$. Initial guesses for the input parameters are provided by the user, and thereafter refined by the fitting routine. Any variable, be it one associated with the atmospheric model or not, may be held constant.

Doppler shift of the target, and a polynomial describing wavelength shift are two examples of other ATMOF modules that may provide additional free parameters or be used rigidly.

At present the reference spectrum is the high resolution Solar data described in section 3.1.1 used as a proxy for a standard star spectrum. However, any sufficiently high resolution spectrum may be used as a reference, and a planned future upgrade of the program will allow the reference to be replaced by a fitted model to facilitate forward modelling.

*3.1.4 Instrumental response*

In addition to atmospheric models, customisable instrumental response modules may also be added and fitted with ATMOF. In this instance a very basic primary instrumental response module was employed consisting of a filter function, a scaling factor and a slope. More sophisticated instrumental response functions are likely to be developed with the continued use of the program, and indeed where ATMOF is being developed specifically for the retrieval of mixing ratios from the telluric atmosphere, such functions are already in use (Kenyi *et al.* 2013).

The filter function used is the K-band spectroscopic order blocking filter on Gemini North, it was supplied by the GNIRS team[4], and is now available online[5]; it is very nearly identical to the curve provided by the manufacturer. Where sufficient similar spectra are available, we often augment the filter function with a smoothed residual fitted curve in the manner of Bailey (2013).

A final spectrum generated with ATMOF: including model input spectrum (high resolution solar spectrum), atmospheric model, and basic instrumental response; is finally convolved with an instrumental line shape mode – in this case a Gaussian profile corresponding to the instrumental resolution. The resolution may also be fitted.

---

[4] Gabelle, T., private communication.
[5] http://gemini.edu/sciops/instruments/?q=node/10531 retrieved 13/8/2013

*3.1.5 Determining the state of the atmosphere and instrumental response*

The observed spectrum of the standard star is the product of the real spectrum of the star, the atmospheric transmission and the instrumental response (illustrated in the left hand side of Fig. 3). To determine the atmospheric transmission and instrumental response we require the standard star spectrum to be known. In this work the standard stars are G-type stars, and we assume they have the same spectrum as the Sun. We use the high resolution solar spectrum described in section 3.1.1 as a proxy for the standard star spectrum.

A model observed spectrum is calculated as the product of the standard star proxy with the VSTAR calculated transmission of the Earth's atmosphere (section 3.1.2), and the instrumental response (section 3.1.4). The product is calculated at the full resolution of the standard star proxy, before the effects of the instrumental resolution are applied – the model telluric spectrum is interpolated to the resolution of the stellar proxy. The model observed spectrum is then fitted by ATMOF (section 3.1.3) to the actual observed spectrum, by way of the fitting parameters that describe the model atmosphere and instrumental response.

The final determination of the atmospheric and instrumental response parameters is described by the state of the parameters at the conclusion of fitting.

*3.1.6 Removal of the artefacts of atmosphere and instrumental response from the data*

The final step is removing the affects of atmospheric transmission and instrumental resolution to reveal the planetary spectrum. The planetary data must be divided by the telluric atmospheric model and instrumental response functions as determined by the procedure described in section 3.1.5. This process is illustrated in the right hand side of Fig. 3). It is reasonable to consider the instrumental profile for the planetary observation to be the same as it was for the standard star observation in cases like that presented here, where the slit width is less than the seeing. Fitting the planetary observation rather than the standard star observation, which we do for the purpose of wavelength calibration, gives a similar fitted instrumental resolution.

The telluric atmospheric transmission model must be modified to account for differences in observing zenith angle between the standard star and planetary observation. This is done by changing and using VSTAR to recalculate the transmission spectrum for the different zenith angle.

The VSTAR model of the atmosphere is at a higher resolution than the data. We need to have both at the same resolution for a satisfactory division. Forward modelling the planetary spectrum at the full resolution of the Earth atmosphere model is the best solution to this problem, but is beyond the scope of this work. Simply convolving the atmospheric model with the instrumental resolution can introduce small errors, so we take a more sophisticated three-step approach.

The process is designed to have the atmospheric transmission act in a similar way to how it does in ATMOF fitting. The first step is to multiply the high resolution solar spectrum by the full resolution atmospheric model and then convolve the product using the instrumental resolution, just as is done by ATMOF. The high-resolution solar spectrum is then also convolved to the instrumental resolution, and the final step is to divide the first spectrum by the second to remove the solar spectrum component, and become the atmospheric model used in the remaining calculations.

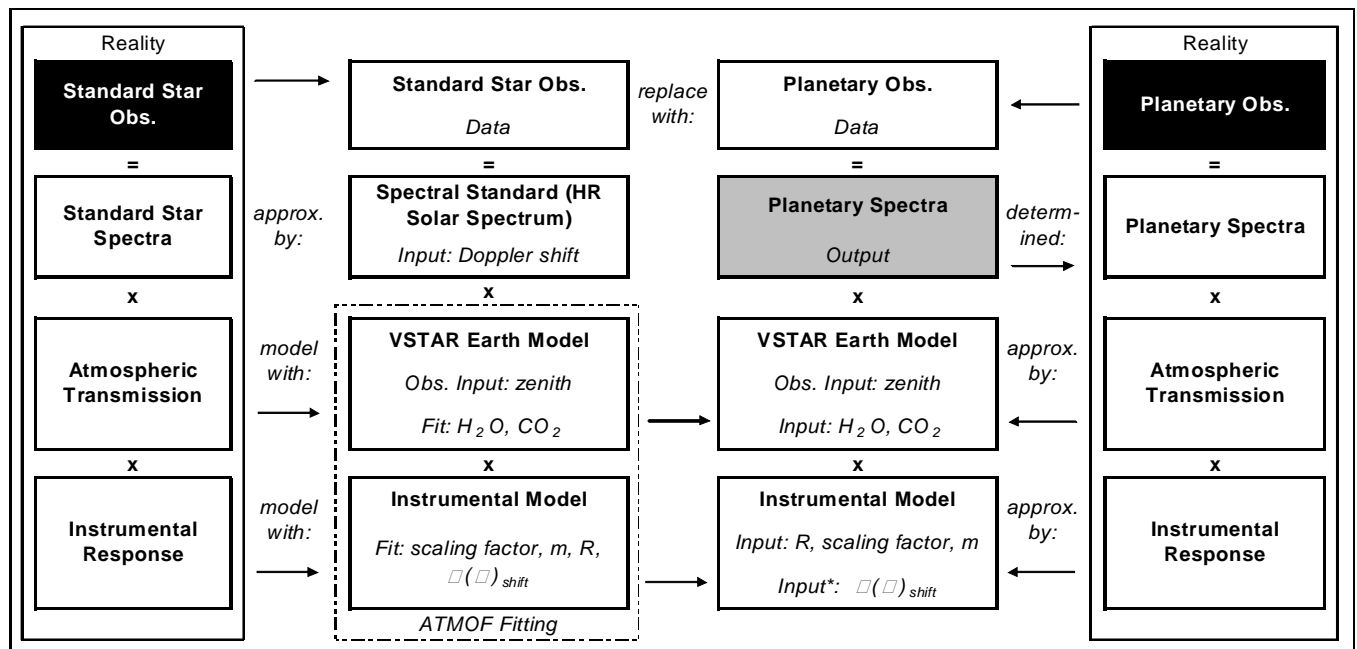

Fig. 3. Modelling approach schematic diagram: the outside columns represent the real processes that produce the observed spectra (black boxes). The second column illustrates how the data from this process is used by ATMOF to determine the state of the atmosphere and the instrumental response. And the third column shows how the outputs from ATMOF are used along with the data from the planetary observation to reveal the real planetary spectra sans any telluric effects. Parameters marked as being fit or as being inputs are representative of typical fitting parameters/inputs; some of these are the zenith angle, the mixing ratios for $H_2O$ and $CO_2$, the scaling factor for the instrumental response, the slope ($m$), the instrumental resolution ($R$), and the wavelength shift ($\lambda(\lambda)_{shift}$). Note that the wavelength shift of the planetary observation labelled as input* in column three is not obtained from the ATMOF fitting in column two; it must be obtained either by another application of ATMOF or through other means.

## 4 EXAMPLES

### 4.1 Titan

*4.1.1 Telluric division*

We used HIP 64345 as the telluric standard calibrator to remove strong absorption due to bands of $CO_2$ in the Earth's atmosphere from Titan's spectrum centred the 2.008 and 2.055 μm (Fig. 4 (a) and (b)). The divided spectrum is shown in panel (c) of Fig. 4. It is clear that residual features due to telluric lines are present in the region around 2.01 μm. Although HIP 64345 is fairly featureless, a few absorption lines are obviously present in a spectral range under consideration and they create spurious emission lines in the Titan spectrum. Only the modelling approach can deal properly with this problem.

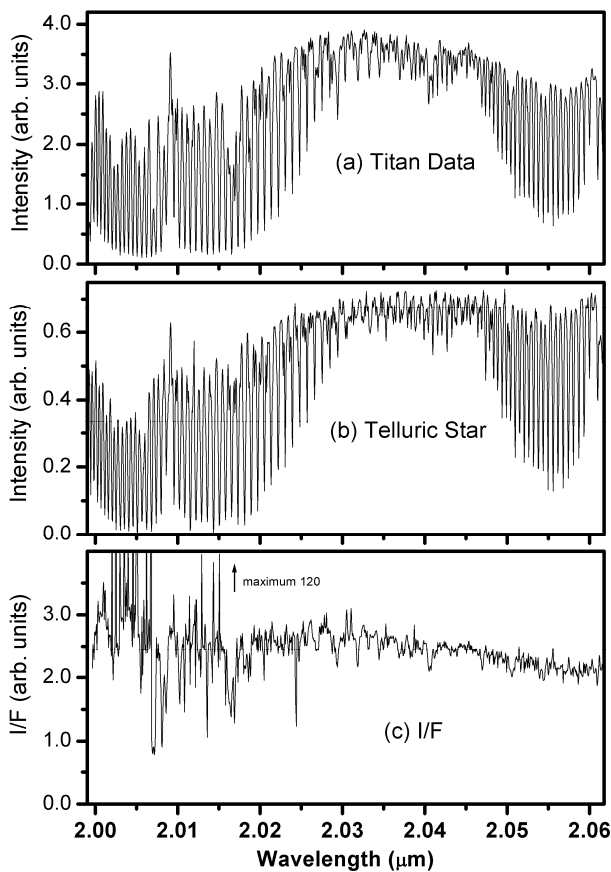

Fig. 4. Titan data processed using the telluric division method.

*4.1.2 Modelling approach*

In this instance two observations of different telluric stars had been made over the same wavelength range with the instrument in the same configuration. Consequently it was possible to augment the GNIRS K-band filter function described in section 3.1.3 in a basic way to help account for otherwise un-parameterised optical components. Each of these telluric star observations was fitted by ATMOF with an atmospheric model. In each instance the zenith angle for the VSTAR model was fixed at the angle of observation whilst the $CO_2$ and $H_2O$ content was fitted. A wavelength shift, Doppler shift, a scaling factor were also free parameters in the fit.

Separately, for each of the two observations the observed spectrum was divided by the model spectrum. The divided spectra were passed through a low-pass Fourier filter to remove any high frequency structure that might be associated with an inaccurate wavelength shift or genuine noise, etc. The result is two augmentation instrumental response functions; the mean of these was taken, and this mean was applied in subsequent ATMOF fitting along with the filter function.

The data for the HIP 63435 observation was then fitted by ATMOF using an instrumental response function incorporating the augmentation function. The fitted parameters were a scaling factor, wavelength shift, instrumental resolution, Doppler shift, $CO_2$ and $H_2O$ mixing ratio. Fig. 5 (a) shows the fit achieved, and (b) the residuals. It is worth remembering here that the stellar proxy is passed through the model atmosphere at a much higher resolution; only after the Gaussian smoothing corresponding to the instrumental resolution does it resemble what is presented in the figure. Thus, here the full benefit of forward modelling is achieved, and there are no introduced errors.

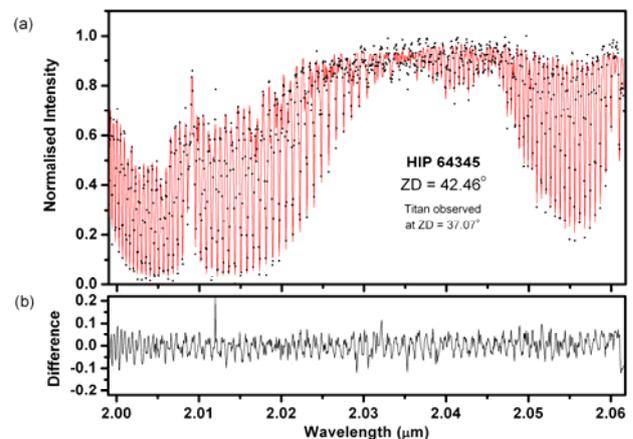

Fig. 5. ATMOF fit of telluric star observation for Titan.

We note at this point that the residuals in Fig. 5 (b) display a regular period with wavelength. This is almost certainly the result of electronic pattern noise. The GNAAC detector controller in GNIRS is known to superimpose vertical banding on the data[6]. There is a routine that is applied as a matter of course to remove it, but its effectiveness is sometimes imperfect. The routine, by default, is set to remove banding 16 pixels in width, which corresponds to 62 cycles over the wavelength range displayed – a very close match. The regular electronic pattern noise is an excellent example of a systematic error neutralised by the modelling approach. Because the entire spectrum is fit the periodic error is cancelled out by the fitting.

The returned fitting parameters: $CO_2$ and $H_2O$ mixing ratio are then used to generate an atmospheric model at the zenith angle of the Titan observation, but prior to this the wavelength shift of the Titan observations needs to be more accurately determined. To achieve this we use ATMOF in an inelegant yet effective way: we replace the telluric star observation data with the Titan observational data. $CO_2$ and $H_2O$ mixing ratios were entered as fixed based on the previous fitting results, the zenith angle of the Titan observation was used. The augmented filter instrumental response previously derived was also used, other parameters were fitted by ATMOF.

By fitting the Titan data rather than the telluric data ATMOF fits the wavelength shift based on the positions of the telluric atmospheric lines – because there are many more of these than there are calibration lamp lines, we get a fine adjustment to the wavelength shift.

---

[6] http://www.gemini.edu/?q=node/11654 retrieved 11/11/2013

The next step is to take the Titan data and divide it by the atmospheric transmission model – using the zenith angle of the observation. However, the VSTAR model of the atmosphere is at a higher resolution than the data. The three step process described in section 3.1.6 is used to produce the atmospheric transmission spectrum from the VSTAR model. The difference between this spectrum and that simply convolved to the instrumental resolution and interpolated to the same grid as the data is negligible, but the procedure is carried out nonetheless.

The Titan observational data (Fig. 6 (a)) is then divided by the atmospheric model (Fig. 6 (b)), and the remaining instrumental response corrections – like the augmented filter response and the scaling factor (as determined in fitting the telluric spectra) (Fig. 6 (c)) – applied to give a spectrum of Titan without the effects of the Earth's atmosphere (Fig. 6 (d)).

The wavelength shift is not constant across the whole spectrum, so in this case a final adjustment was made at the end in a piecewise fashion. Plotting the final spectrum with a progression of wavelength shifts around the fit determined value highlights local deviations. An inaccurate wavelength shift looks similar to an incorrect Doppler shift – around a spectral line a sharp spike and trough will be present as a pair. Visual inspection revealed three distinct regions (likely to be related to calibration arc-line positions): < 2.0094 μm, 2.0094 – 2.0167 μm, > 2.0167 μm. Total shifts applied in these regions were: -0.20 Å, -0.02 Å, and 0.04 Å respectively. Adjustments to the wavelength shift in these regions were made on the basis of the progression of wavelength shift plots. Even such small shifts as these can have a significant effect at this resolution.

To provide a comparison with the telluric division method we divide the spectra produced (I) by the Solar flux (F) (Fig. 6 (e)). This gives the radiance factor (I/F) which is the standard way of representing the reflected light from a solar system object, and can be directly compared with models. The same high resolution solar spectrum used in the atmospheric fit is also used for this purpose. By dividing by the Solar flux the stellar light reflected from the planet can be removed from the spectrum, but only if the Doppler shift is correct. This I/F plot is shown in Fig. 6 (f).

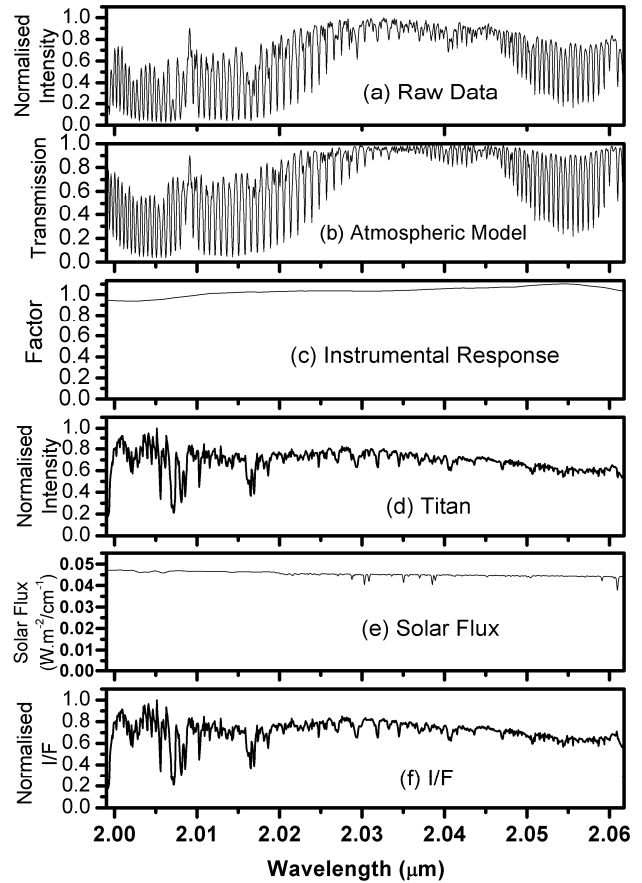

Fig. 6. Key spectra in the processing of Titan data using the modelling approach.

*4.1.3 Comparison*

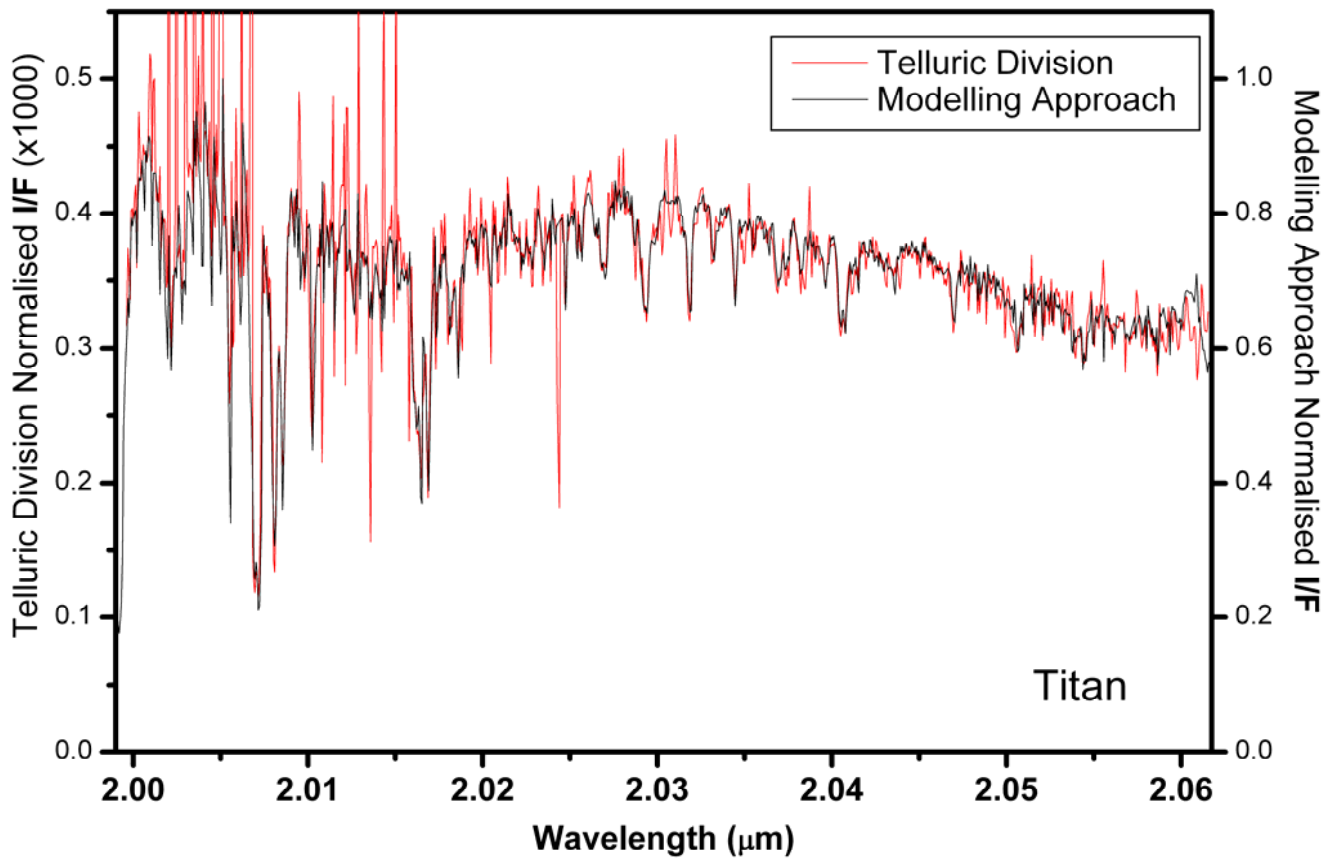

Fig. 7. Comparison of telluric division and modelling approach for Titan observation (I/F).

Fig. 7 shows the difference between the I/F spectra obtained for Titan for the two methods. Overall the spectra display the same trends, indicating that both methods are doing fundamentally the same thing. The most obvious difference between the two is seen between 2 – 2.02 μm. Large spikes and troughs in the telluric division spectrum are the result of noise. In this region atmospheric absorption significantly reduces the signal to noise. Whilst the telluric division method divides one noisy spectrum by another here, the modelling approach uses an atmospheric model that is derived from every data point in the spectrum of the telluric star. Though not as great an effect, between ~2.05 – 2.06 μm atmospheric absorption also reduces signal to noise; here the telluric division spectrum is also clearly noisier, with unphysical positive spikes present.

If one considers Fig. 6 (e) and notes the positions of prominent solar lines at, for example, ~2.03 μm and ~2.06 μm peak-trough pairs associated with the Doppler shift of the telluric star with respect to the reflected Solar light from Titan's atmosphere are also present in the telluric division spectra, but not in the modelling approach spectra. The modelling approach therefore has clear advantages in this region.

The only spurious features are these due to uncorrected Solar absorption lines, which are not properly corrected in division by the solar type standard star due to its different Doppler shift.

**4.2 Jupiter**

*4.2.1 Telluric division*

A traditional approach of removing telluric absorption due to the Earth's atmosphere worked better for the spectrum of Jupiter than Titan. Better observing conditions and wavelength calibration as discussed in section 2.3 gave a clean output spectrum in Fig 8(c).

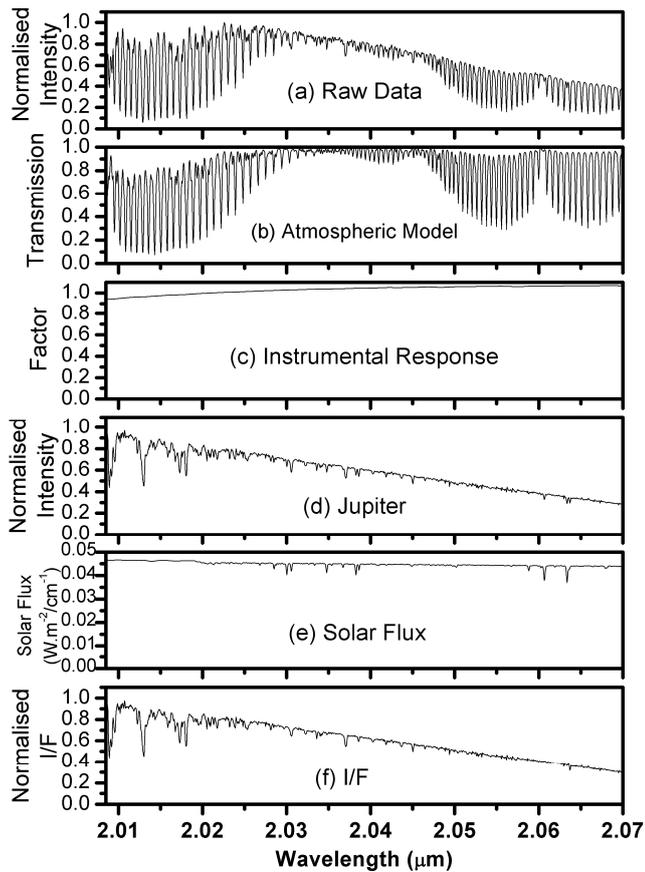

Fig. 8. Jupiter data processed using the telluric division method.

*4.2.2 Modelling Approach*

Only one appropriate telluric star observation was available for the Jupiter observation, so no augmented response was calculated. The first step here was fitting the telluric star observation.

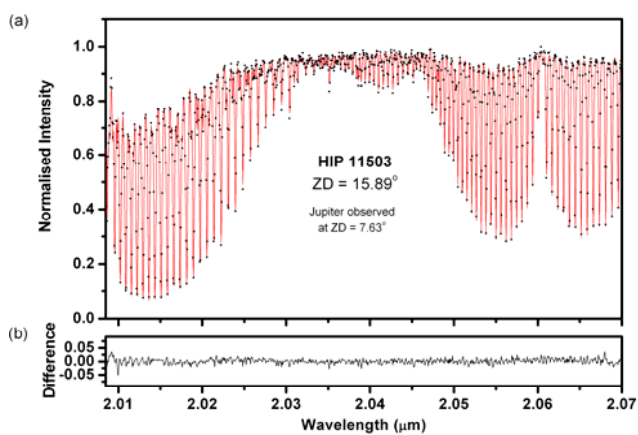

Fig. 9. ATMOF fit of telluric star observation for Jupiter.

The data for the HIP 11503 observation was fitted by ATMOF using the following fitting parameters: scaling factor, linear slope, instrumental resolution, Doppler shift, $CO_2$ and $H_2O$ mixing ratio. Here we fit a quadratic function to describe wavelength shift. Fig. 9 (a) shows the fit achieved, and (b) the residuals.

Examination of Fig. 9 (b) reveals the influence the replaced part of the high resolution solar data has on the fit. Instead of a low level noise signature in this region, the fluctuations in the residuals are coarser. The fitted atmospheric model spectrum is minimally affected here, because the fit includes a significant portion of high resolution data in the remainder of the spectrum.

The RMS for the residuals in the units of Fig. 9 (b) is ~0.018. The S/N varies from ~50 for the deepest telluric lines, to ~300 for the regions of the spectrum with the least telluric absorption; the greatest contributors to the formal error are photon shot noise and detector read noise, the sum of these ranges from ~0.0015 to ~0.0032 respectively. Other contributors to the mismatch between model and observation include noise in the Kitt Peak solar spectrum, any unremoved contribution from electronic pattern noise, the fact that HIP11503 won't have the exact spectrum of the Sun as assumed; and probably the greatest contributor will be inaccuracies in the line lists used to generate the model.

As in the Titan case, we next fitted the Jupiter observation with ATMOF directly to obtain the wavelength shift. $CO_2$ and $H_2O$ mixing ratios were entered as fixed based on the previous fitting results, the zenith angle of the Jupiter observation was used. Other parameters were fitted by ATMOF, including the three coefficients of a quadratic fit to the wavelength shift.

Fig. 10 shows the improvement ATMOF's fitting of the wavelength shift has over the calibration made using the arc lamp lines. There is a difference of between 0.15 and 0.4 Angstrom, more importantly, the ATMOF determined values match much closer those determined by visual inspection – a series of constant wavelength shifts were applied to the data, and the value best corresponding to elimination of a peak/trough pair selected for a number of prominent lines. Our ability to discern the effects of an imprecise wavelength shift using this method to such a high degree of agreement with the software highlights the importance of this improvement.

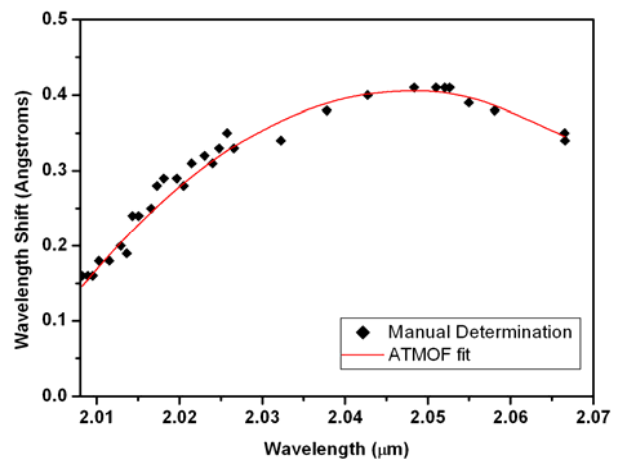

Fig. 10. The quadratic fit of the wavelength shift achieved by ATMOF. The shift is relative to the calibration made with the arc lamp. The data points represent the determination made upon visual inspection.

To calculate the model atmosphere to divide the Jupiter observational data with, the same three step procedure used for Titan and described in section 3.1.6 was followed. Again, the difference between the spectrum so produced and one simply smoothed to the instrumental resolution was negligible. The VSTAR Earth atmosphere model used the Jupiter observational zenith angle, the $CO_2$ and $H_2O$ mixing ratios determined from the fit of the telluric star, and the quadratic wavelength shift determined as described above. The result is shown as Fig. 11 (b), this was then divided by the instrumental response (Fig. 11 (c)). The instrumental response incorporates the filter function – described in section 3.1.4 – as well as the scaling factor and linear slope found from fitting the telluric star.

scattered light depends on instrument optics and is wavelength dependant.

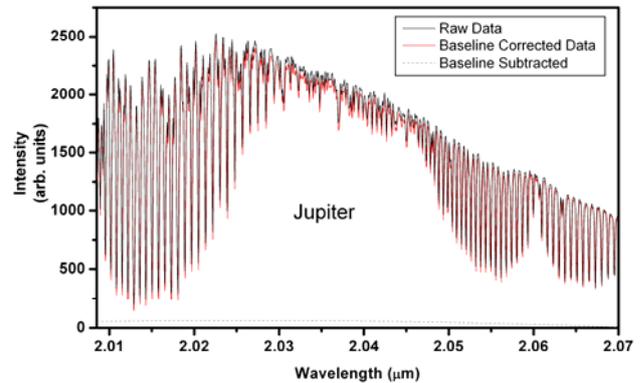

Fig. 12. Jupiter data before and after baseline correction.

In order to make a correction to the baseline we assumed that there was no significant $CO_2$ absorption in Jupiter's atmosphere. The accepted $CO_2$ mixing ratio for Jupiter is low enough that this is a good assumption (Lellouch *et al.* 2002). This allowed us to produce a series of Jupiter spectra, each with a different subtraction to the baseline. Which spectra corresponded to a flattening of the telluric $CO_2$ absorption lines was determined by inspection at three different wavelengths. A quadratic fit of these three points was then used to take account of the wavelength dependence of spurious scattered light. The baseline correction is shown in Fig. 12, and the final form of the data used in Fig. 11 (a).

Following the adjustment for baseline correction the Jupiter spectrum shown in Fig. 11 (d) was produced, and then divided by the appropriately Doppler shifted Solar flux (Fig. 11 (e)) to produce the I/F plot in Fig. 11 (f) for comparison with the telluric method.

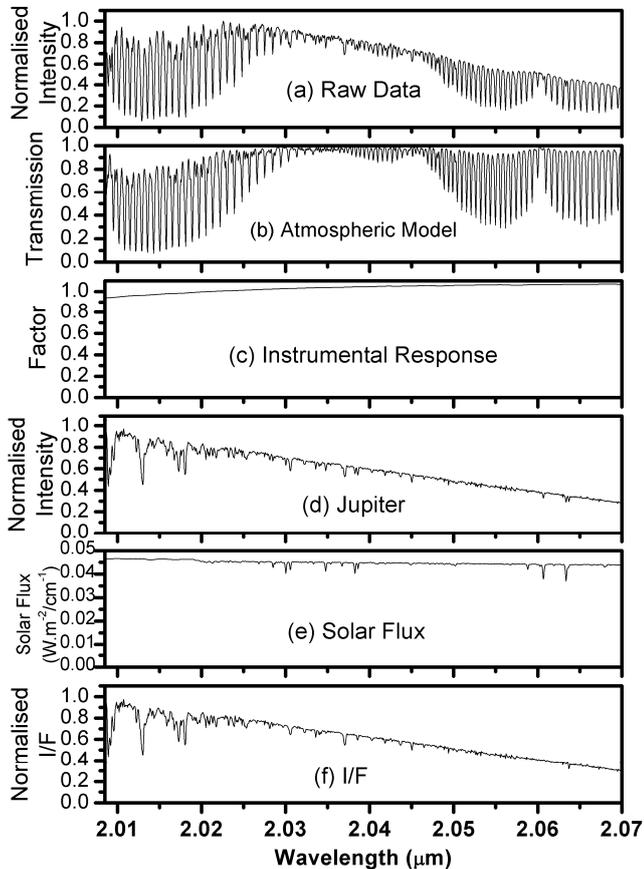

Fig. 11. Data reduction using the modelling approach for Jupiter observations.

At this point, on the first run through, an unexpected challenge was encountered. Differences in the baseline subtraction procedure between the telluric star observation and the Jupiter observation meant the division created spurious pseudo-emission-like lines in the final spectrum. These lines correspond, predominantly, to the positions of $CO_2$ and other absorption lines in the Earth's atmosphere.

Ordinarily an object is observed in two locations in the telescope slit, and an "ABBA" procedure followed where the background light in the slit is subtracted as the baseline. This was done for the telluric star observations, and the Titan observation. However, Jupiter fills the whole slit and a different procedure had to be used: instead a blank sky frame was subtracted as the baseline. This didn't properly account for all the scattered light from the planet though, and the background was underestimated. Correcting for this is not straight-forward because the amount of

*4.2.3 Comparison*

The Jupiter example presented here is just about the ideal set of circumstances for the telluric division method. Both the planetary and standard star observation were made at high elevation, meaning that the atmosphere through the line of sight is very similar. Jupiter is very big and bright – giving a good signal to noise ratio even in low transmission regions of the spectrum. Humidity was low at the time of observation as well, meaning the influence of telluric water on the spectrum was small (incidentally, ATMOF fitted a lower $H_2O$ mixing ratio for Jupiter than Titan).

Despite these advantages the modelling approach is demonstrably better. In Fig. 13 the most noticeable advantage of the modelling approach is the lack of prominent peak-trough pairs corresponding to the Doppler shifted Solar/stellar lines. A reduction in the noise in the region from 2.01 to 2.025 μm and from ~2.05 μm onward is also noticeable. No matter how good the signal is from the target object, the modelling approach will always have less noise associated with it, because the Earth atmosphere model is the result of fitting the whole of the telluric star spectrum.

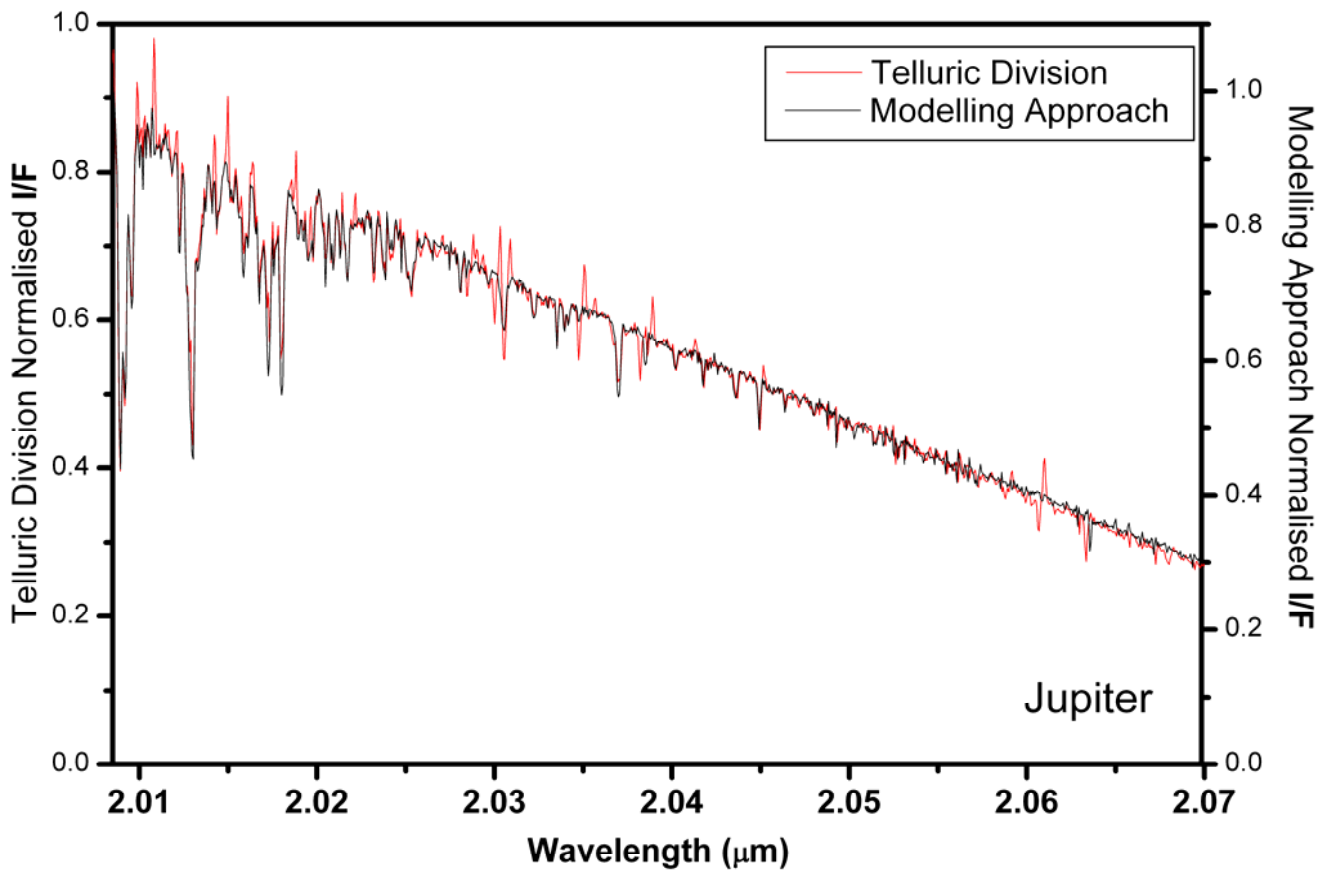

Fig. 13. Comparison of telluric division and modelling approach for Jupiter observation (I/F).

Less obvious in the final comparison, but of no less importance, the wavelength calibration is significantly improved. The initial arc-lamp calibration applied has been further refined by ATMOF using the telluric absorption lines. Because the calibration has been carried out independently for the telluric star and planetary observations the division of the Jupiter data by the atmospheric model is more precise than for the telluric division technique, and it is calibrated to the very precise telluric line positions from our model.

## 5 DISCUSSION AND CONCLUSIONS

Here we have demonstrated how a modelling approach has clear and demonstrable advantages over the standard technique of telluric division. The most apparent improvements are: (1) a markedly reduced influence of noise in low transmission regions of the Earth's atmosphere, (2) the introduction of no spurious features from a difference in Doppler shift of the standard star from reflected solar light, and (3) an improved wavelength calibration.

The modelling method is much less dependent on the choice of standard star. It is largely immune to differences in zenith angle and spectral type that could cause problems with the telluric division method. Noise and instrumental artefacts in the standard star observation, such as the electronic pattern noise discussed in section 4.1.2, have minimal effect on the final resulting spectrum.

It is particularly noteworthy that even given ideal conditions during our observations of Jupiter – low humidity, planetary and telluric star observations both near zenith – the modelling approach gave a visibly less noisy spectrum without the artefacts inherent to telluric division.

The improved confidence we have in the wavelength calibration allows for easier identification of absorption lines. Comparison of the spectra of Jupiter and Titan in Fig. 14 reveals their different characteristics. Titan in this region has absorption features due to $CH_3D$ (Bailey *et al.* 2011) that are potentially useful for determining its D/H ratio. Jupiter does not appear to have the same features, but has absorption due to $CH_4$ and $NH_3$.

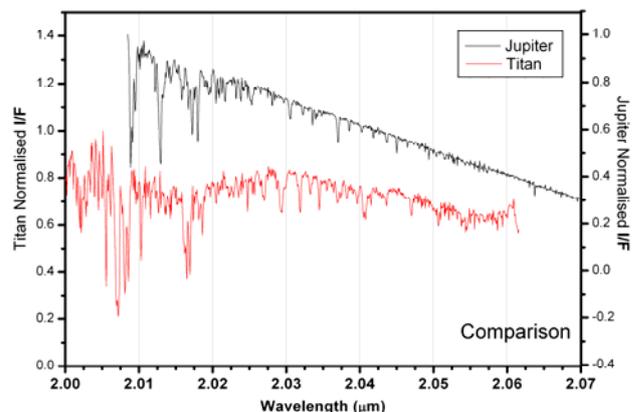

Fig. 14. A comparison of I/F for the Titan and Jupiter data processed in this work. Note that neither data set is flux calibrated, and so the intensity scales have been chosen purely to display the data on the same plot.

The modelling approach demonstrated here is a significant improvement on telluric division for removing telluric absorption;

its benefits are greatest where molecular absorption lines near saturation and linear corrections for air mass are inappropriate. Our method eliminates the problems inherent in dividing regions of the spectrum with a 'forest' of fine lines by a low resolution standard where we fit the telluric star data. This falls short of the power of a full forward modelling approach, as our telluric model must still be smoothed to a lower resolution to retrieve the planetary data from the telluric contaminated planetary data. A full forward model avoids this limitation where the spectra of the object of interest can be well parameterised and modelled. However, ATMOF has been designed so that with further development it can be used for full forward modelling. ATMOF's modular design means that the stellar proxy we used may be easily substituted. With some alteration it will be able to be substituted for a fitted model. And so the modelling approach here represents not only the realisation of one mode of operation, but also an important stepping stone in the program's development of a second mode.

Finally, it is worth saying something about the versatility of the program, and this approach. We have here demonstrated its effectiveness for a small part of the infrared K-band when observing solar system bodies with atmospheres. The same procedures and techniques may be applied effectively in any spectral region where transmission is attenuated without being completely saturated by telluric absorption lines, so long as the high resolution spectrum of a standard is well know in that region. Clearly, high resolution data in strongly attenuated parts of the Earth's atmosphere is uncommon, but high resolution (static) models may be substituted. Further, this technique has further applications within our solar system in the observation of icy moons, or beyond it in the observation of brown dwarfs and exo-planets in particular.

**ACKNOWLEDGEMENTS**

The work presented here is based on observations obtained at the Gemini Observatory (programs GN-2011A-Q-36 and GN-2011B-Q-1), which is operated by the Association of Universities for Research in Astronomy, Inc., under a cooperative agreement with the NSF on behalf of the Gemini partnership: the National Science Foundation (United States), the National Research Council (Canada), CONICYT (Chile), the Australian Research Council (Australia), Ministério da Ciência, Tecnologia e Inovação (Brazil) and Ministerio de Ciencia, Tecnología e Innovación Productiva (Argentina). The work was supported by the Australian Research Council through Discovery grant DP 110103167.

**REFERENCES**

Allen D.A., Cragg T.A., 1983, MNRAS, 203, 777
Bailer-Jones C.A.L., Lamm M., 2003, MNRAS, 339, 477
Bailey J., 2013, AMTD, 6, 1067
Bailey J., Ahlsved L., Meadows V.S., 2011, Icar, 218, 218
Bailey J., Kedziora-Chudczer L., 2012, MNRAS, 419, 1913
Bailey J., Simpson A., Crisp D., 2007, PASP, 119, 228
Beer A., 1852, Ann Phys u Chem, 86, 78
Bouchet P., Schmider F.X., Manfroid J., 1991, A&AS, 91, 409
Buemann W., Linter B.R., Koven C.D., Angert A., Pinzon J.E., Tucker C.J., Fung I.Y., 2007, PNAS, 104, 4249
Carter B.S., 1990, MNRAS, 242, 1
Chamberlain S.A., Bailey J.A., Crisp D., 2006, PASA, 23, 119
Encrenaz T., 2008, SSRv, 135, 11
Kedziora-Chudczer L., Bailey J., Horner J., 2013, in: Short W., Cairns I. eds., Proceedings of the 12th Australian Space Science Conference, National Space Society of Australia, p. 53
Kenyi C., Cotton D.V., Bailey J., 2013, in: Short W., Cairns I. eds., Proceedings of the 12th Australian Space Science Conference, National Space Society of Australia, p. 117
Lallement R., Bertin P., Chassefière E., Scott N., 1993, A&A, 271, 734
Lellouch E., Bézard B., Moses J.I., Davis G.R., Drossart P., Feuchtgruber H., Bergin E.A., 2002, Icar, 159, 112
Lundström I., Ardeberg A., Maurice E., Lindgren H., 1991, A&AS, 91, 199
McGregor P.J., 1994, PASP, 106, 508
Press W.H., Teukolsky S.A., Vetterling W.T., Flannery B.P., 1992, *Numerical Recipies in FORTRAN: The Art of Scientific Computing*, Cambridge University Press,
Raupach M.R., Marland G., Ciais P., Le Quéré C., Candell J.G., Klepper G., Field C.B., 2007, PNAS, 104, 10288
Rothman L.S., et al., 2009, JQSRT, 110, 533
Seifahrt A., Käufl H.U., Zängl G., Bean J.L., Richter M.J., Siebenmorgen R., 2010, A&A, 524, 1
Swain M.R., et al., 2009, Nat, 463, 637
Tinetti G., et al., 2007, Nat, 448, 169
Wallace L., Livingston W., Hinkle K., Bernath P., 1996, ApJ, 106, 165
Widemann T., Bertaux J.-L., Querci M., Querci F., 1994, A&A, 282, 879